\documentclass[fleqn,usenatbib]{mnras}

\usepackage{newtxtext,newtxmath}

\usepackage[T1]{fontenc}
\usepackage{ae,aecompl}

\usepackage{graphicx}	
\usepackage{amsmath}	
\usepackage{amssymb}	

\usepackage{xcolor}

\title[Chromosome Maps of M54]{Light element variations within the different age-metallicity populations in the nucleus of the Sagittarius dwarf}

\author[A. Sills et al.]{Alison Sills,$^{1}$\thanks{E-mail: asills@mcmaster.ca (AS)}
Emanuele Dalessandro,$^{2}$
Mario Cadelano,$^{2,3}$
Mayte Alfaro-Cuello$^{4}$
\newauthor{and J.~M.~Diederik Kruijssen$^{5}$}
\\
$^{1}$Department of Physics \& Astronomy, McMaster University, 1280 Main Street West, Hamilton, ON, L8S 4M1, Canada\\
$^{2}$INAF - Astrophysics and Space Science Observatory Bologna, Via Gobetti 93/3, 40129, Bologna, Italy\\
$^{3}$Dipartimento di Fisica e Astronomia, Universit\`a degli Studi di Bologna, Via Gobetti 93/2, 40129, Bologna, Italy\\
$^{4}$Max-Planck-Institut f\"ur Astronomie, K\"onigstuhl 17, 69117 Heidelberg, Germany\\
$^{5}$Astronomisches Rechen-Institut, Zentrum f\"ur Astronomie der Universit\"at Heidelberg, M\"onchhofstra{\ss}e 12-14, 69120 Heidelberg, Germany
}
\date{Accepted 2019 September 26. Received 2019 September 26; in original form 2019 August 9}

\pubyear{2019}

\begin{document}
\label{firstpage}
\pagerange{\pageref{firstpage}--\pageref{lastpage}}
\maketitle

\begin{abstract}
The cluster M54 lies at the centre of the Sagittarius dwarf spheroidal galaxy, and therefore may be the closest example of a nuclear star cluster. Either in-situ star formation, inspiralling globular clusters, or a combination have been invoked to explain the wide variety of stellar sub-populations in nuclear star clusters. Globular clusters are known to exhibit light element variations, which can be identified using the photometric construct called a chromosome map. In this letter, we create chromosome maps for three distinct age-metallicity sub-populations in the vicinity of M54. We find that the old, metal-poor population shows the signature of light element variations, while the young and intermediate-age metal rich populations do not. We conclude that the nucleus of Sagittarius formed through a combination of in-situ star formation and globular cluster accretion. This letter demonstrates that properly constructed chromosome maps of iron-complex globular clusters can provide insight into the formation locations of the different stellar populations. 
\end{abstract}

\begin{keywords}
globular clusters: general -- globular clusters: individual: M54 -- galaxies:dwarf
\end{keywords}



\section{Introduction}

Massive star clusters in the Milky Way and other nearby galaxies show signatures of star-to-star variations in light elements such as C, N, O, Na, while iron is typically constant within the observational errors \citep[see the recent review by][]{BL18}.  The source of the light element abundance variations is unknown, but is likely tied to helium production in hot hydrogen burning, and is tightly connected to the cluster environment. 

A few globular clusters do show star-to-star iron variations. Two clusters with the largest variations are M54, which is known to be at the centre of the Sagittarius dwarf galaxy \citep[e.g.][]{Siegel07}; and $\omega$ Centauri which has long been hypothesized to be the former nucleus of an accreted dwarf \citep[e.g.][]{BekkiFreeman2003}. Indeed,  the similarity between $\omega$ Centauri and the M54/Sagittarius system has been discussed in the literature \citep{Carretta2010wCen}. Nuclear star clusters are found in many galaxies \citep{GeorgievBoker2014}, but their formation mechanism is still debated. Two main ideas exist in the literature. Either the stars in the nuclear star cluster formed {\it in situ} in the centre of the host galaxy \citep[e.g.][]{Schinnerer2008}, or nuclear star clusters are the amalgamation of infalling globular clusters \citep[e.g.][]{Gnedin2014}. Lately, there has been some evidence that both processes may be at work to build up the stellar populations of nuclear star clusters \citep[e.g.][]{Hartmann2011}. It is tempting to make the connection between nuclear star clusters in other galaxies, and the iron-complex globular clusters in our own Galaxy, particular as we can study the local globular clusters in much more detail. 

Recent MUSE IFU observations of M54 \citep{Mayte, Mayte2} have uncovered a very complex star formation history and kinematic separation of the stellar populations in the region. The populations can be divided into three main groups: an old, metal-poor component (OMP) older than about 10 Gyr but with a fairly large age spread, and a mean metallicity of [Fe/H]=-1.4 that makes up the bulk of the stars; an intermediate, metal-rich component (IMR) with a broad age range of 3-9 Gyr and a mean metallicity of [Fe/H]=-0.3; and a young, metal-rich component (YMR) with ages younger than about 3 Gyr (a mean age of $\approx$ 2 Gyr) and a mean metallicity of [Fe/H]=-0.04. These three populations are not only clearly separated in age-metallicity space, but also have distinct spatial and kinematic properties. The interpretation of \citet{Mayte} is that the OMP population represents globular clusters that fell in to form the nucleus in the past. They suggest that there are sufficiently large age and metallicity spreads within this population that it may have been more than one cluster. They postulate that the YMR population was formed {\it in situ}, possibly triggered by a passage of the Sagittarius dwarf galaxy in the disk of the Milky Way \citep[see also][]{TepperGarcia2018}. The IMR population shares some characteristics with stars in the bulk of the galaxy, and so these could be Sagittarius field stars, which may or may not be bound to M54. Earlier work \citep{Bellazzini2008, Carretta2010, Mucciarelli17} used slightly different nomenclature: they called only the metal-poor population ``M54" and concluded it was a separate entity from the more metal-rich Sagittarius nucleus, which they considered to be a single group of stars. 

The traditional way to investigate stellar populations in detail is to collect spectra of a few giants, since spectroscopy of such faint stars is expensive and difficult, particularly in crowded regions. There has been some spectroscopic work on light element abundances in stars in Sagittarius \citep{Carretta2010}. They studied 76 giants in from the metal-poor giant branch of M54, and 27 giants from the more metal-rich population. Signs of the sodium-oxygen anti-correlation were found within the metal-poor objects, but not in the metal-rich stars. While this work is indicative, we would like to be able to study a much larger number of stars from all three populations within M54. Fortunately, recently a photometric method for probing light element abundances has been developed \citet{Milone2017}. The so-called `chromosome map' capitalizes on the sensitivity of blue HST filters to nitrogen abundances, and a very long wavelength baseline to probe temperature and therefore helium abundance in giants. With a creative combination of 4 HST filters, a distinctive pattern emerges, and we can separate large numbers of stars into `enriched' and `normal' populations. By applying this technique to the various age-metallicity populations in M54, we can look for signatures of light element variations. 

We would expect that the OMP population should show strong evidence of the light element variations indicative of a globular cluster origin, possibly with more complexity if this population consists of more than one globular cluster. The IMR stars should have constant light element abundances, consistent with their metallicities and formation in a normal part of the Sagittarius dwarf galaxy. The YMR stars could show light element variations if the formation environment in a nuclear star cluster is similar to that in a globular cluster, or will show no light element variations if their formation region is more like the field, or possibly a low-mass cluster.

In this letter, we explore whether the chromosome maps of the different populations in the iron-complex cluster M54 can be used to determine if any of these previously-identified age-metallicity populations show clear signs of multiple populations, and therefore formation in a clustered environment.
A similar analysis has been done for the chromosome map of $\omega$ Centauri \citep{Milone2017, Marino2019}, although they did not use metallicities and ages to separate their populations but used only the different giant branch morphologies. A spectroscopic study of the unusual cluster Terzan 5 suggests that its two different iron populations have different aluminum and oxygen abundances, but perhaps not with the same patterns as in normal globular clusters \citep{origlia11}, and unfortunately this cluster is behind sufficient amounts of extinction that it is unlikely that we can create a chromosome map soon. The complexity of M54 falls between that of $\omega$ Centauri and Terzan 5 on one hand, and some of the other iron-spread clusters (such as M22, M2, and NGC 1851) on the other. We wish to determine whether we can use photometric methods to disentangle the formation site of stars in these less extreme clusters. 

\section{Methods}

\begin{figure}
\begin{center}
\includegraphics[width=\columnwidth]{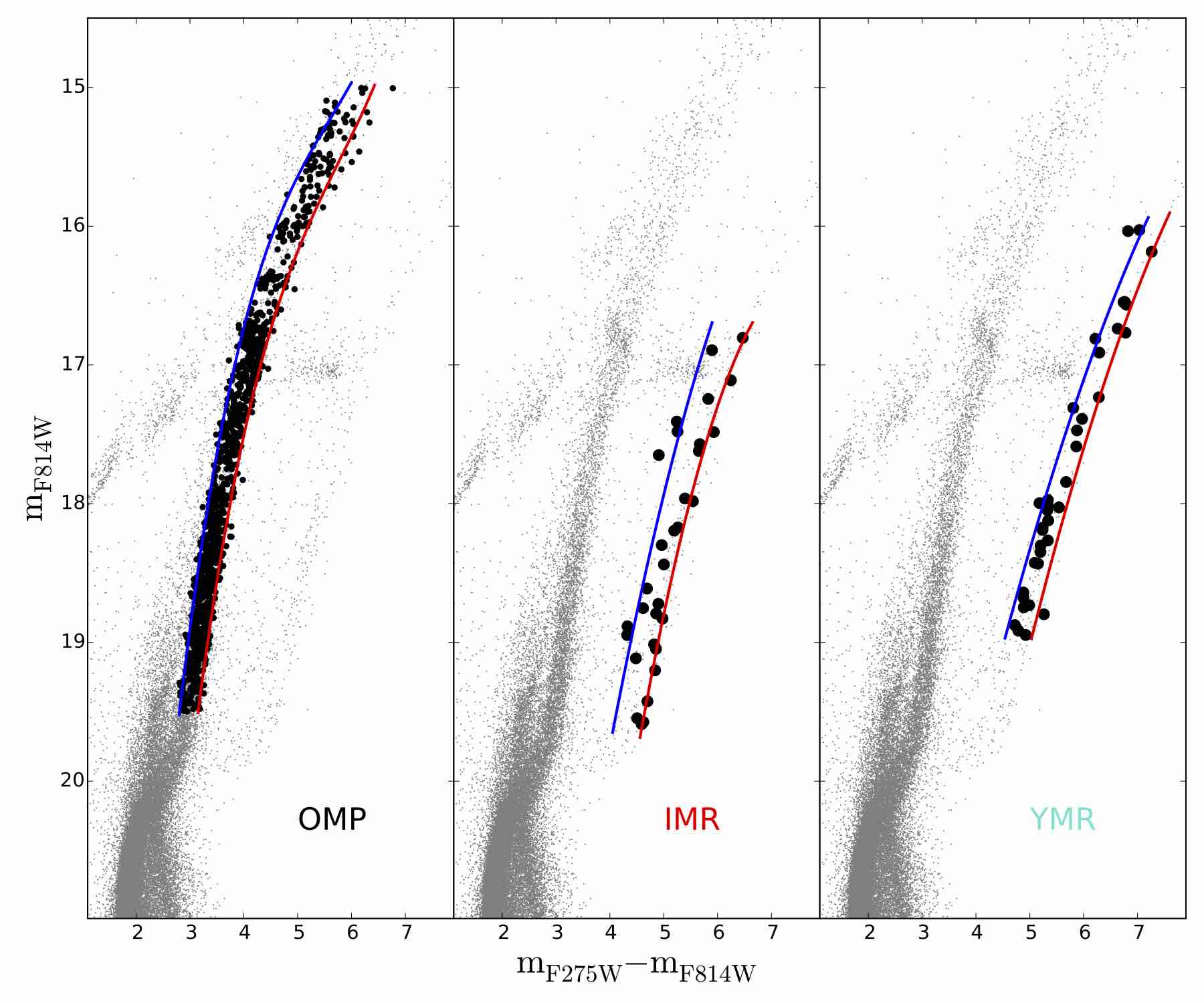}
\caption{F275W-F814W colour-magnitude diagrams for all stars (in grey) with each population highlighted with large black points in their own panel, as labelled. The fiducial lines for each population are shown. \label{fig:CMD}}
\end{center}
\end{figure}

\begin{figure}
\begin{center}
\includegraphics[width=\columnwidth]{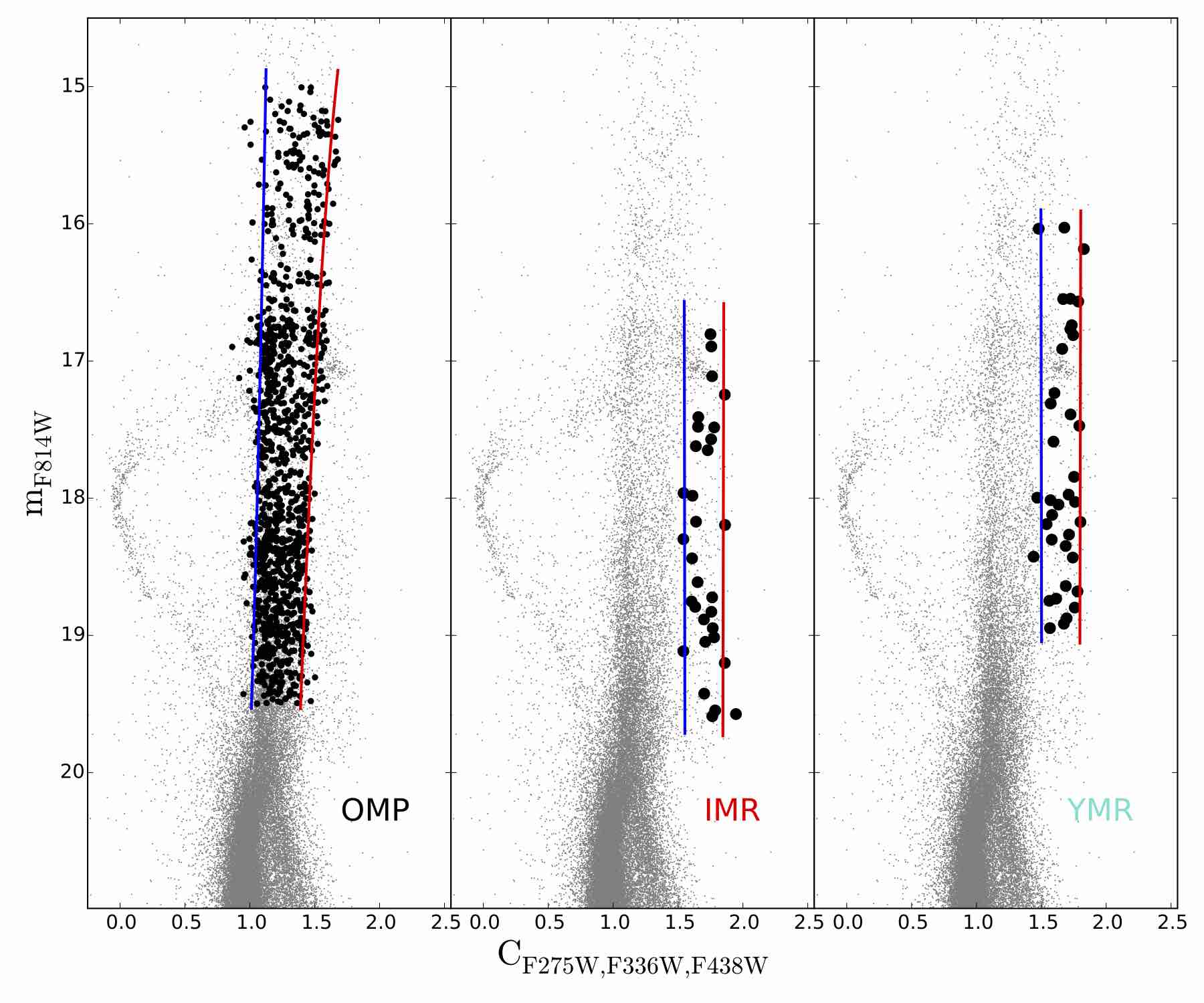}
\caption{F275W,F336W,F438W pseudo-colour-magnitude diagrams for all stars (in grey) with each population highlighted with large black points in their own panel, as labelled. The fiducial lines for each population are shown. \label{fig:CUBI}}
\end{center}
\end{figure}

We cross-matched the catalogue used by \citet{Mayte} to determine the age-metallicity populations in M54 with the HST UV Globular Cluster Survey data \citep{Nardiello2018, Piotto2015} to determine their F275W, F336W, F475W, and F814W magnitudes. We chose the `method 1' photometric catalogue of \citet{Nardiello2018}, although their other two methods provided almost identical results for the giants in M54. Following the method outlined in \citet{Milone2017}, we created separate chromosome maps for the three populations. We split the sample of stars in common in the groups: OMP component, which includes stars with -2.5<[Fe/H]<-1.5, IMR including stars with -1.5<[Fe/H]<-0.3 and YMR, [Fe/H]>-0.3. Briefly, for each group we derived two fiducial lines both in the (F275W-F814W, F814W) and ((F275W-F336W)-(F336W-F438W), F814W) colour-magnitude diagrams which are the 5th and 95th percentiles of RGB star distribution. We then verticalised the distribution of RGB stars and normalised them for the intrinsic RGB width ($W$) at 2 mag brighter than the turnoff in the F814W band, thus deriving $\Delta_{F275W,F814W}$ and $\Delta_{F275W,F336W,F438W}$. The (F275W-F814W, F814W) colour-magnitude diagrams of the three components are shown in Figure \ref{fig:CMD}, clearly showing the effect of metallicity on the giant branch morphology. The pseudo-colour-magnitude diagram is shown in Figure \ref{fig:CUBI}. We selected fiducial RGB stars with 15<F814W<19.5,16<F814W<19.5 and 16<F814W<19 for OMP, IMR and YMR respectively. We also selected stars with the quality parameter $qfit>0.9$ in the F275W, F336W, F438W and F814W bands and with a membership probability $>75\%$ \citep[see][for details]{Nardiello2018}. Table \ref{tab:table} gives the number of giants we identified in each population, along with the giant branch widths in each (pseudo)colour-magnitude diagram used to create the chromosome maps.

\begin{table}
\centering
\caption{Number of stars and giant branch widths for each population.}
\label{tab:table}
\begin{tabular}{cccc}
\hline
Population & N & $W_{275,814}$ & $W_{275,336,438}$\\
\hline
OMP & 1108 & 0.39 & 0.40 \\
IMR & 31 & 0.45 & 0.30 \\
YMR & 37 & 0.33 & 0.30 \\
\end{tabular}
\end{table}

 To directly compare the derived chromosome maps, we need to take into account their dependence on metallicity as shown by \citet{Marino2019}. The shape of the chromosome map indeed depends on the metallicity of the cluster. The various patterns are driven primarily by the effects of helium and nitrogen on the different colours used to create the map, and the same absolute change in either of those elements will cause a greater or lesser change in colours depending on the underlying amount of the elements in all stars in the cluster. Fortunately, \citet{Marino2019} characterized the shape differences, and provides metallicity-dependent corrections to create what they call a `universal' chromosome map. We applied those corrections to our three populations, using the average metallicity of the stars within each population, to obtain $\delta_{F275W,F814W}$ and $\delta_{F275W,F336W,F438W}$.

\section{Results}

\begin{figure}
\begin{center}
\includegraphics[width=\columnwidth]{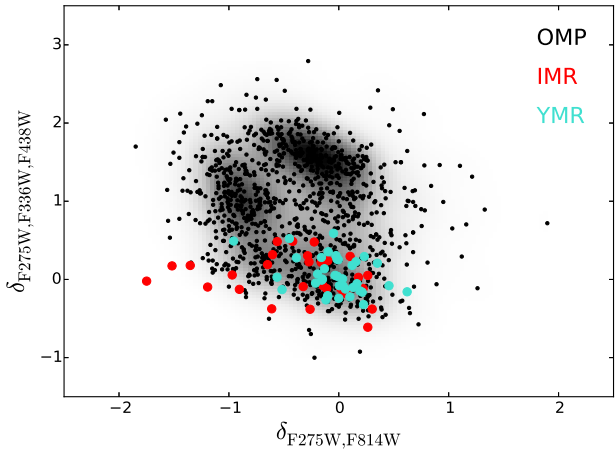}
\caption{Metallicity-normalized chromosome map. The old, metal-poor population is shown in black and shows the clear signature of `multiple populations' with three main concentrations in this diagram. The intermediate age, metal-rich population (IMR, in red) and the young, metal-rich population (YMR, in blue), however, belong only to the the lower right clump, which is where stars with `normal' or field composition are expected to lie.  \label{fig:chromomap}}
\end{center}
\end{figure}

In Figure \ref{fig:chromomap} we show the metallicity-independent chromosome map for stars in the three populations in the central region of Sagittarius. The OMP stars clearly split into at least three regions in the diagram, confirming our expectation of a globular cluster origin for these stars.  The IMR and YMR populations, however, are both clustered in the bottom right of the diagram, which is where field stars or stars with `normal' composition are expected to lie. 

We wish to emphasize the importance of using the metallicity normalization of \citet{Marino2019} when comparing chromosome maps of different populations. We looked at our analysis at each stage of constructing the maps, and only with the metallicity normalization did we see this clear separation of the populations. When we used a single fiducial line for the giant branch and a single width, as was done in \citet{Milone2017}, the cluster nature of the OMP was apparent but both the IMR and YMR stars were scattered throughout the plane. Going to individual fiducial lines and widths, to start to take into account the metallicity effects on the giant branch shape, was an improvement but even then, the two metal-rich populations still had a wide range in values, particularly in the $\Delta_{F275W,F336W,F438W}$ axis. The correct analysis, including the metallicity correction, is required to fully exploit the power of the chromosome maps. 

We note that the number of stars in each of the IMR and YMR populations is small, and therefore the placement of the fiducial lines is more uncertain than for the OMP population. We estimated the uncertainty in the $W$ values given in table 2, and confirmed that the location of the IMR and YMR stars in the chromosome map are not significantly changed by considering the possible ranges of $W$ in either axis. Note that the IMR and YMR populations are much younger than the globular clusters used by \citet{Marino2019} to define their metallicity correction. It is therefore not clear if such a correction is valid also for these young populations. However, since metallicity is the main parameter affecting giant branch colours, and its effect goes at all ages in the same direction, we consider such a correction a reasonable approximation.

\section{Summary and Discussion}

We have shown that the three different age-metallicity populations identified by \citet{Mayte} in M54 have different photometric characteristics as shown in the chromosome map plane. We interpret these results to say that the OMP population has a globular cluster-like origin, but the IMR and YMR populations do not. Our conclusions are consistent with the spectroscopic results of \citet{Carretta2010}. The YMR population is seen to be centrally concentrated within the Sagittarius dwarf, even more so than the OMP population \citep{Mayte}, but it may not have sufficient mass to show the light element variations seen in the more massive globulars. The results presented here suggest that the nucleus of Sagittarius may have been formed through a combination of in-situ star formation and globular cluster accretion. On the basis of this chromosome map, we do not formally distinguish between the accretion of a single globular cluster, or multiple globular clusters. We note, however, that the chromosome map of the OMP stars retains significant complexity, unlike the chromosome maps of simpler clusters like NGC 6838 and NGC 6397 \citep[][their figure 12]{Marino2019}. Further investigation of this population is warranted.

We conclude that it is possible to use chromosome maps to determine the formation site of different stars in iron-complex globular clusters. With improved telescope facilities, it may be possible to use this technique to probe nuclear star clusters in nearby galaxies. A number of ingredients are required: high-precision photometry in 4 bands (3 in the UV/blue region of the spectrum and one quite red band); low-resolution spectroscopy to divide the populations into different metallicity groups; and a careful analysis of the resulting CMDs that includes both unique fiducial lines and giant branch widths for each population, plus the metallicity-dependent correction to the chromosome map axes. 

We note that in our analysis we have used the average metallicity for each subpopulation to correct the chromosome maps, but according to the analysis of \citet{Mayte}, the metallicity spread within a population is 0.1-0.2 dex. It is possible that an even more fine-grained analysis, particularly within the OMP population, may be able to give us tighter groupings in the chromosome map plane. It would be interesting to see if we could use a more sophisticated version of this method to determine if nuclear star clusters were indeed formed through the merger of more than one globular cluster. 

Metallicity changes the giant branch morphology, but age can also play a role. It would be interesting to do an analysis like that of \citet{Marino2019} but with age instead of metallicity. Almost all the globular clusters in the HST survey are quite old, but evidence of light element variations have been seen in clusters down to 2 Gyr \citep{martocchia18} so it would be worth characterizing the chromosome maps for younger populations. 

We intend to apply this method to many of the other iron-complex clusters that have been identified through their chromosome maps or through spectroscopy. With the advent of the MUSE globular cluster survey \citep{kamann18}, we can now determine metallicities of many thousands of globular cluster stars, and clearly separate the different populations within these iron-complex clusters. The chromosome map, particularly in its `universal' form, can be a useful tool to attempt to disentangle the possible formation histories of these unusual clusters.

\section*{Acknowledgements}

We thank the Sexten Centre for Astrophysics for providing a beautiful venue for fruitful scientific discussions. AS and JMDK gratefully acknowledge funding from Sonderforschungsbereich SFB 881 ``The Milky Way System'' (subproject B2) of the German Research Foundation (DFG). JMDK gratefully acknowledges funding from the DFG in the form of an Emmy Noether Research Group (grant number KR4801/1-1) and from the European Research Council (ERC) under the European Union's Horizon 2020 research and innovation programme via the ERC Starting Grant MUSTANG (grant agreement number 714907). AS is supported by the Natural Sciences and Engineering Research Council of Canada. 




\bibliographystyle{mnras}
\bibliography{M54.bib} 

\bsp	
\label{lastpage}
\end{document}